\documentclass[review,number,sort&compress]{elsarticle}
\usepackage{graphicx}
\usepackage{amssymb}
\usepackage{color,soul}
\usepackage{url}
\usepackage{ulem}
\usepackage{wasysym}
\usepackage{subcaption}
\usepackage[figuresright]{rotating}

\definecolor{orange}{rgb}{1,0.5,0}
\definecolor{cadmiumgreen}{rgb}{0.0, 0.42, 0.24}

\journal{Applied Radiation and Isotopes}

\begin{document}

\begin{frontmatter}


\title{A comparison of untagged gamma-ray and tagged-neutron yields 
from $^{241}$AmBe and $^{238}$PuBe sources}


\author[lund,ess]{J.~Scherzinger}
\author[glasgow]{R.~Al~Jebali}
\author[glasgow]{J.R.M.~Annand}
\author[lund,ess]{K.G.~Fissum\corref{cor1}}
\ead{kevin.fissum@nuclear.lu.se}
\author[ess,midswe]{R.~Hall-Wilton}
\author[lund]{S.~Koufigar\fnref{fn2}}
\author[lund]{N.~Mauritzson}
\author[lund,ess]{F.~Messi}
\author[lund,ess]{H.~Perrey}
\author[lund]{E.~Rofors}

\address[lund]{Division of Nuclear Physics, Lund University, SE-221 00 Lund, Sweden}
\address[ess]{Detector Group, European Spallation Source ESS AB, SE-221 00 Lund, Sweden}
\address[glasgow]{SUPA School of Physics and Astronomy, University of Glasgow, Glasgow G12 8QQ, Scotland, UK}
\address[midswe]{Mid-Sweden University, SE-851 70 Sundsvall, Sweden}

\cortext[cor1]{Corresponding author. Telephone:  +46 46 222 9677; Fax:  +46 46 222 4709}
\fntext[fn2]{present address: Applied Physics and Applied Mathematics Department, Columbia University, New York, NY 10027, USA}

\begin{abstract}
Untagged gamma-ray and tagged-neutron yields from $^{241}$AmBe and 
$^{238}$PuBe mixed-field sources have been measured. Gamma-ray spectroscopy
measurements from 1 -- 5 MeV were performed in an open environment using a 
CeBr$_3$ detector and the same experimental conditions for both sources. 
The shapes of the distributions are very similar and agree well with 
previous data. Tagged-neutron measurements from 2 -- 6 MeV were performed in 
a shielded environment using a NE-213 liquid-scintillator detector for the 
neutrons and a YAP(Ce) detector to tag the 4.44~MeV gamma-rays associated 
with the de-excitation of the first-excited state of $^{12}$C. Again, the 
same experimental conditions were used for both sources. The shapes of these
distributions are also very similar and agree well with previous data, each
other, and the ISO recommendation. Our $^{238}$PuBe source provides 
approximately 2.6 times more 4.44~MeV gamma-rays and 2.4 times more neutrons 
over the tagged-neutron energy range, the latter in reasonable agreement 
with the original full-spectrum source-calibration measurements performed 
at the time of their acquisition.
\end{abstract}


\end{frontmatter}

\section{Introduction}
\label{section:introduction}

Actinide/Be-based radioactive sources are typically used for cost-effective 
fast-neutron irradations. Neutrons are produced when the $\alpha$-particle 
from the decay of the actinide interacts with the $^{9}$Be nucleus. If one
is only interested in neutrons, a major 
drawback associated with these sources is the gamma-ray field produced by 
the original actinide. Higher-energy gamma-rays are also produced via the
$\alpha$ + $^{9}$Be $\rightarrow$ n + $^{12}$C$^{*}$ reaction. However, if 
detected,
these 4.44~MeV gamma-rays can be used to ``tag" the corresponding 
neutrons~\cite{scherzinger15}, resulting in a polychromatic energy-tagged 
neutron beam. In this paper, we employ the neutron-tagging technique to 
measure tagged, fast-neutron yields from $^{241}$AmBe and $^{238}$PuBe using
a NE-213 liquid-scintillator detector. We also measure the corresponding
untagged gamma-ray yields using a Cerium-bromide (CeBr$_3$) detector. Our
goal was to measure the tagged-neutron yields provided by the two sources 
and indentify which provided the higher tagged-neutron yield.

\section{Apparatus}
\label{section:apparatus}

\subsection{Actinide/Be-based sources}
\label{subsection:be_source}

For the investigations performed in this work, both $^{241}$Am/$^{9}$Be 
(Am/Be) and $^{238}$Pu/$^{9}$Be (Pu/Be) sources were employed. Both
actinides decay predominantly via the emission of $\alpha$-particles.
According to NuDat~\cite{nudat}, $^{241}$Am decays via the emission of 25 
different
$\alpha$-particles (5.4786~MeV weighted-mean energy) while $^{238}$Pu 
decays via the emission of 14 different $\alpha$-particles (5.4891~MeV
weighted-mean energy). The difference in the weighted mean 
$\alpha$-particle energies is thus only about 10~keV. When these 
$\alpha$-particles interact with $^{9}$Be, fast neutrons are produced. 
Because the weighted mean of the incident $\alpha$-particle energies is
essentially the same for both actinides, the energy spectra of emitted 
neutrons are anticipated to demonstrate strong similarities.
These neutrons have a 
maximum energy of about about 11~MeV~\cite{lorch73}. If the recoiling 
$^{12}$C is left in its first-excited state, the freed neutron is 
accompanied by an isotropically radiated prompt 4.44~MeV de-excitation 
gamma-ray. The radiation fields associated with these sources are thus
a combination of the gamma-ray field associated with the original actinide 
together with 4.44~MeV gamma-rays and their associated fast-neutrons.
Our Am/Be source radiates approximately 1.14~$\times$~10$^6$ neutrons 
per second~\cite{natphyslab}, while our Pu/Be source radiates 
approximately 2.99~$\times$~10$^6$ neutrons per 
second~\cite{radiochemicalcentre}, both nearly isotropically.

\subsection{Detectors}

\begin{figure} 
\begin{center}
\resizebox{9.5cm}{7cm}{\includegraphics{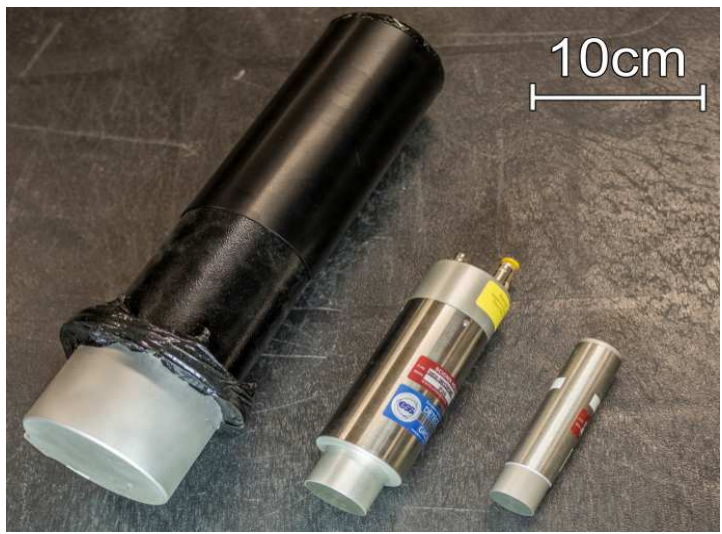}}
\caption{The detectors employed for the measurements presented in this
paper. From the left: NE-213 liquid-scintillator neutron and gamma-ray 
detector; CeBr$_3$ gamma-ray spectroscopy detector; YAP(Ce) gamma-ray 
trigger detector.}
\label{figure:detectors}
\end{center}
\end{figure}

The NE-213 liquid-scintillator and YAP(Ce) and CeBr$_3$ gamma-ray 
detectors used in the measurements for this work are shown 
in Fig.~\ref{figure:detectors}.

\subsubsection{NE-213 liquid-scintillator detector}
\label{subsubsection:ne213_detector}

NE-213~\cite{ne213} is a standard organic liquid scintillator which has been
employed for several decades
for detecting neutrons in strong gamma-ray fields. Because gamma-ray induced
scintillations in NE-213 are generally fast (10s of ns decay times) while 
neutron induced scintillations are much slower (100s of ns decay times), the 
type of radiation incident upon the NE-213 scintillator may be identified 
by examining the time structure of the scintillation pulses. The 
NE-213 liquid-scintillator detector employed in this 
measurement
consisted of a 3~mm thick cylindrical aluminum cell (62~mm long $\times$ 
94~mm $\diameter$) coated internally with EJ-520 TiO$_2$-based reflective 
paint~\cite{ej520} which contained the NE-213. A 5~mm thick borosilicate 
glass plate~\cite{borosilicate} was used as an optical window. The filled 
cell was coupled to a cylindrical PMMA UVT lightguide~\cite{pmma} 
(57~mm long $\times$ 72.5~mm $\diameter$) coated on the outside by 
EJ-510~\cite{ej510} TiO$_2$-based reflective paint. The assembly was 
coupled to a $\mu$-metal shielded 7.62~cm ET Enterprises 9821KB 
photomultiplier tube (PMT) and base~\cite{et_9821kb}. Operating voltage 
was set at about $-$1900 V, and the energy calibration was determined 
using standard gamma-ray sources together with a slightly modified 
version of the method of Knox and Miller~\cite{knox72}. See 
Refs.~\cite{scherzinger15,scherzinger16} for more detail.

\subsubsection{YAP(Ce) 4.44~MeV gamma-ray detectors}
\label{subsubsection:yaps}

The YAP(Ce) (YAlO$_{3}$, Ce$^{+}$ doped) gamma-ray detectors 
employed in this measurement were provided by Scionix~\cite{scionix}.
Each detector was comprised of a cylindrical (2.54~cm long $\times$
2.54~cm $\diameter$) crystal~\cite{moszynski98} attached to 2.54~cm
Hamamatsu Type R1924 PMT~\cite{hamamatsu}. YAP(Ce) is a useful 
gamma-ray trigger scintillator in a strong radiation field as it is 
both radiation hard and relatively insensitive to fast neutrons.
Operating voltage was set at about $-$800~V, and energy calibration 
was determined using standard gamma-ray sources. These YAP(Ce) 
detectors were used to count the 4.44~MeV gamma-rays eminating from 
the sources and thus tag the corresponding emitted neutrons. See 
Refs.~\cite{scherzinger15,scherzinger16} for more detail.

\subsubsection{CeBr$_3$ gamma-ray detector}
\label{subsubsection:cebr3}

The CeBr$_3$ gamma-ray detector employed in this measurement was provided by 
Scionix~\cite{scionix}. The detector was comprised of a cylindrical (3.81~cm 
long $\times$ 3.81~cm $\diameter$) crystal~\cite{billnert11} attached to a
5.08~cm Hamamatsu Type R6231 PMT~\cite{hamamatsu}. Due to its fast response 
without slow scintillation components and excellent energy resolution 
(3.8\% FWHM for the full-energy peak produced by the 662~keV gamma-ray from
$^{137}$Cs), CeBr$_3$ is a useful scintillator for 
gamma-ray spectroscopy. Operating voltage was set at about $-$850~V, and 
energy calibration was determined using standard gamma-ray sources. The 
CeBr$_3$ gamma-ray detector was used to measure the gamma-ray spectra
associated with the Am/Be and Pu/Be sources.

\section{Measurement}
\label{section:measurement}

\subsection{Setup}
\label{subsection:setup}

The measurements described below were performed sequentially for each of the 
actinide/Be sources. The CeBr$_3$ detector used for gamma-ray spectroscopy was 
located 50~cm from the source and placed at source height. Neutron shielding
by organic materials or water was not employed to avoid production of 2.23~MeV 
gamma-rays from neutron capture on protons. The threshold for the CeBr$_3$ 
detector was set at about 350~keV. For the neutron-tagging measurements, water 
and plastic shielding were employed to define a neutron beam. Four YAP(Ce) 
4.44~MeV gamma-ray trigger detectors were individually positioned around the 
source at a distance of about 10~cm and placed slightly above the height of 
the source. The threshold for the YAP(Ce) detectors was also set at about 
350~keV. The NE-213 detector was located 110~cm from the source and placed 
also at source height. The threshold for the NE-213 detector was set at about 
200~keV$_{ee}$ (keV electron equivalent). In the energy region above 1~MeV, all
of the detectors triggered largely on the 4.44~MeV gamma-rays emanating from 
the source. The NE-213 detector triggered on both gamma-rays and neutrons 
produced by the source.  Neutron time-of-flight (TOF)
and thus energy was determined by detecting both the fast neutron in the 
NE-213 detector and the prompt correlated 4.44~MeV gamma-ray in a YAP(Ce) 
detector. By tagging the neutrons in this fashion, the neutron yield as a 
function of kinetic energy was measured. Note that due to the energy lost to 
the 4.44~MeV gamma-ray, the tagging technique restricted the maximum available 
tagged-neutron energies to about 7~MeV.

\subsection{Electronics and data acquisition}
\label{subsection:electronics_and_daq}

The analog signals from the NE-213 detector were sent to Phillips Scientific 
(PS) 715 NIM constant-fraction timing discriminators (CFDs) as well as LeCroy 
(LRS) 10-bit 2249A (DC-coupled 60~ns short gate SG) and 11-bit 2249W 
(AC-coupled 500~ns long gate LG) CAMAC charge-to-digital converters (QDCs). 
The analog signals from the CeBr$_3$ detector were sent to PS715 NIM CFDs as 
well as CAEN V792 12-bit (DC-coupled 60~ns gate) VME QDCs. For both the 
tagged-neutron and untagged gamma-ray measurements, the CFD signals were used 
to trigger the data-acquisition (DAQ). For the tagged-neutron measurements, the
CFD signals from the NE-213 detector also provided start signals for LRS 2228A 
CAMAC time-to-digital converters (TDCs) used for the neutron TOF determination.
The YAP(Ce) detector provided the corresponding (delayed) stop signal for this
TOF TDC.  A CES 8210 branch driver was employed to connect the \mbox{CAMAC} 
electronics to a VMEbus and a SIS 3100 PCI-VME bus adapter was used to connect 
the VMEbus to a LINUX PC-based DAQ system. The signals were recorded and 
processed using ROOT-based software~\cite{root}. See 
Refs.~\cite{scherzinger15,scherzinger16} for more detail.
 
\section{Results}
\label{section:results}

\subsection{Pulse-shape discrimination (PSD)}
\label{subsection:pulse_shape_discrimination}

\begin{figure} 
\begin{center}
\resizebox{10.5cm}{6cm}{\hspace{-0.5cm}\includegraphics{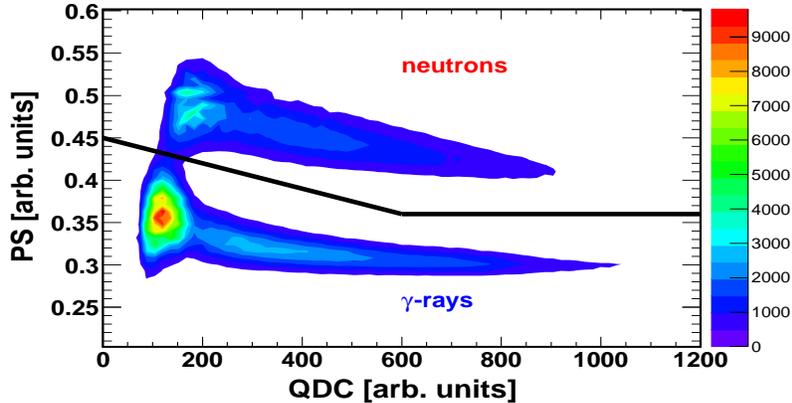}}
\caption{\label{figure:psd}
Typical PSD contour plot obtained using the actinide/Be sources for tagged 
events. PS has been plotted against the total energy deposited in the LG~QDC. 
The upper band corresponds to neutrons while the lower band corresponds to 
gamma-rays. The black line indicates the neutron cut applied to obtain 
subsequent spectra. (For interpretation of the references to color in this 
figure caption, the reader is referred to the web version of this article.)
}
\end{center}
\end{figure}

We employed the ``tail-to-total" 
method~\cite{jhingan08,lavagno10,pawelczak13} to analyze the time
dependence of the scintillation pulses to discern between gamma-rays
and neutrons. The pulse shape (PS) was defined as 
\begin{equation}
PS = (LG - SG) / LG,
\end{equation}
where $LG$ and $SG$ were the integrated charges produced by the
scintillation-light pulses in the LG and SG QDCs, respectively.
Figure~\ref{figure:psd} shows a contour PSD distribution obtained using the 
Pu/Be source for tagged events when the NE-213 detector started the TOF TDC 
and the YAP(Ce) detector stopped the TOF TDC. Tagged background was clearly 
sparse, and the separation between neutrons and gamma-rays was excellent.

\subsection{Time-of-flight (TOF)}
\label{subsection:time_of_flight}

\begin{figure} 
\begin{center}
\resizebox{12.5cm}{15cm}{\hspace{-0.5cm}\includegraphics{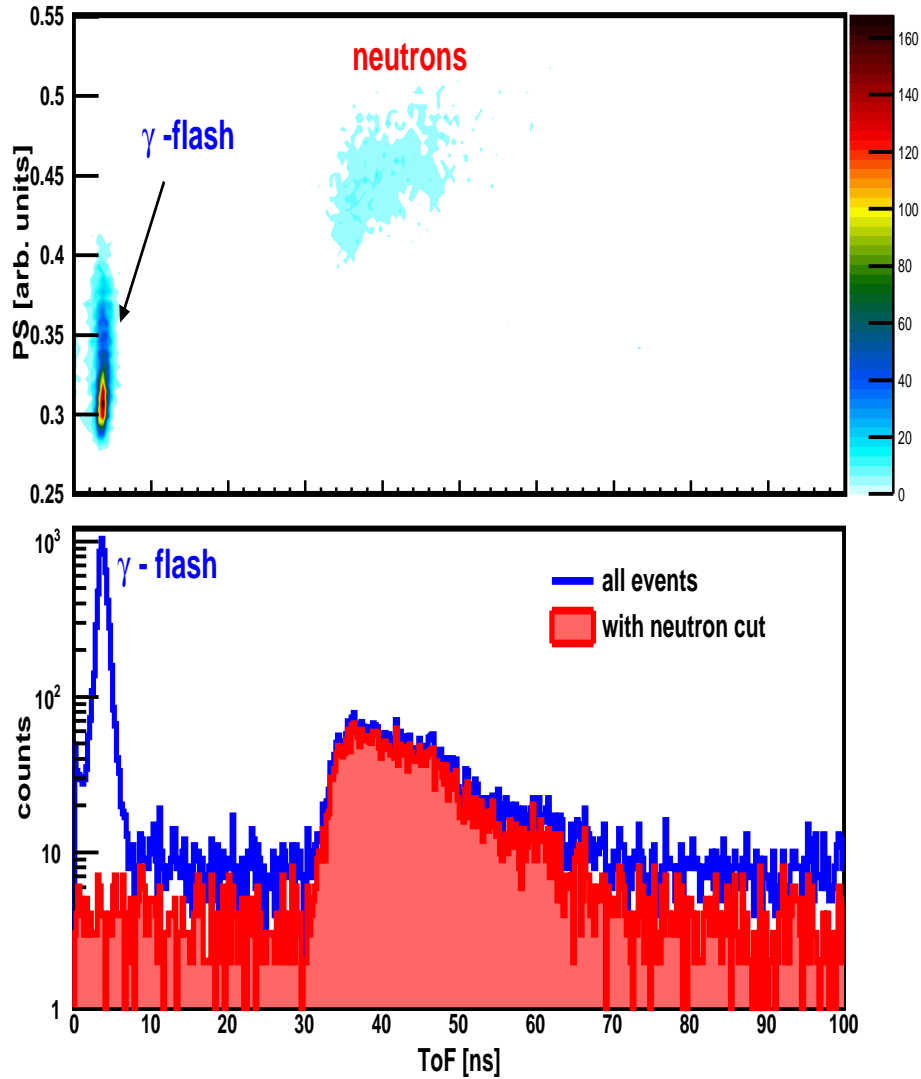}}
\caption{\label{figure:tof}
Typical TOF distributions obtained using the actinide/Be sources. 
In the top panel, PS has been plotted against TOF. Gamma-rays from
the gamma-flash are shown to the left of the plot, while neutrons 
are shown in the center.  In the bottom panel, TOF distributions 
have been plotted for all events (unshaded blue) as well as events 
identified as neutrons (shaded red). The peak in the unshaded blue 
distribution located at about 3.5~ns is the gamma-flash.
(For interpretation of the references to color in this figure caption, 
the reader is referred to the web version of this article.)
}
\end{center}
\end{figure}

Data from the TOF TDC were calibrated and used to establish the
neutron time-of-flight. Correlated gamma-ray pairs originating
in the sources and detected one in the NE-213 detector and one in
a YAP(Ce) detector provided a
``gamma-flash"\footnote{For example, from the $\alpha$ decay of
$^{241}$Am to an excited state of $^{237}$Np.}. $T_{0}$, the
instant of emission of the neutron from the source, was determined
from the location of the gamma-flash in the TOF spectra using the
speed of light and measurements of the distances between the
YAP(Ce) detector, the NE-213 detector, and the source.
In the top panel of Fig.~\ref{figure:tof}, the excellent separation
between gamma-rays and neutrons is illustrated in a contour plot
of PS against TOF. In the bottom panel, the contour plot has been
projected onto the TOF axis. Further, the black-line cut from
Fig.~\ref{figure:psd} has been applied, and a TOF distribution 
for events identified as neutrons is shown. The gamma-flash has
a FWHM of about 1.5~ns (arising from the timing jitter in our
signals) and is located at 3.5~ns. Tagged neutrons dominate the 
plot between 30~ns and 70~ns. Events between 8 and 30~ns and for 
TOF $>$ 70~ns are mainly due to random coincidences. Random events included 
source-related gamma-rays and neutrons and depended on the singles rates in 
the YAP and NE-213 detectors.

\subsection{Untagged gamma-ray yields}
\label{subsection:gamma_ray_yields}

\begin{figure} 
\begin{center}
\resizebox{12.5cm}{8.9cm}{\hspace{-0.5cm}\includegraphics{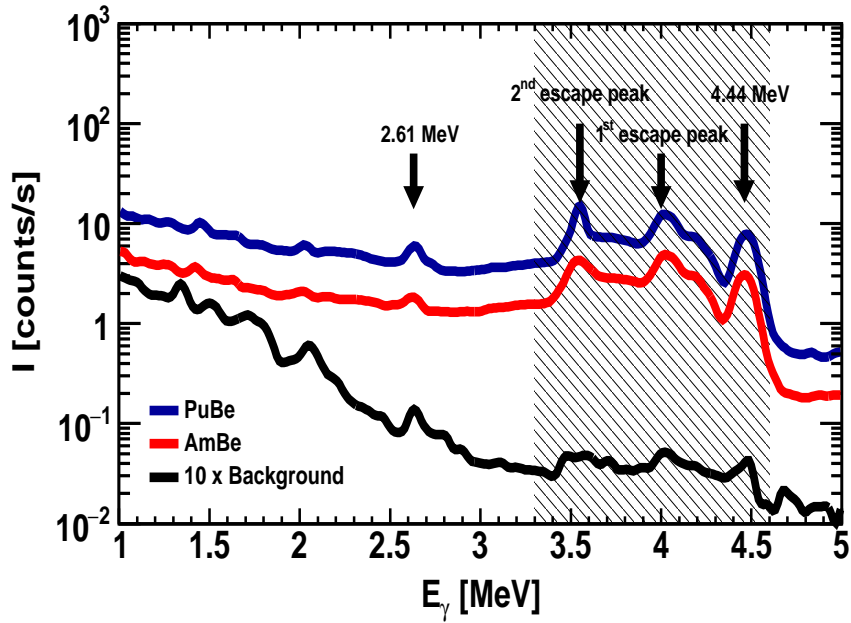}}
\caption{\label{figure:gamma_ray_yield}
Untagged livetime-corrected gamma-ray yields. The dark blue distribution 
corresponds to the gamma-ray field measured from the Pu/Be source, the 
red distribution corresponds to the gamma-ray field measured from the 
Am/Be source, and the black distribution corresponds to the gamma-ray 
field measured without a source.  The shading shows the region of 
integration used for comparing 4.44~MeV gamma-ray yields. (For 
interpretation of the references to color in this figure caption,
the reader is referred to the web version of this article.)
}
\end{center}
\end{figure}

Figure~\ref{figure:gamma_ray_yield} shows our untagged gamma-ray yields 
obtained with the CeBr$_3$ detector. Other than cycling the sources, no 
changes were made to the apparatus. A software cut has been placed at 
1~MeV to exclude very low energy gamma-ray events, including
room-associated background.  Over this energy range, the gamma-ray field 
from the Pu/Be source is clearly stronger than that from the Am/Be 
source, but the structure is very similar. The structures in the Pu/Be 
and Am/Be gamma-ray distributions between about 3 and 5~MeV correspond 
to neutron-associated gamma-rays from the de-excitation of 
$^{12}$C$^{*}$. The full-energy peak appears at 4.44~MeV, while the 
corresponding first- and second-escape peaks appear at 3.93~MeV and 
3.42~MeV, respectively. The average of the ratio of the integrals of 
these yields between 3.3 and 4.5~MeV is 2.6. The peak at 2.61~MeV 
corresponds to the de-excitation of $^{208}$Pb$^{*}$ to its ground state 
and is largely due to isotopic impurities in our sources, and to a lesser 
extent room background.

\subsection{Tagged-neutron yields}
\label{subsection:neutron_yields}

\begin{figure} 
\begin{center}
\resizebox{13.0cm}{12.0cm}{\hspace{-0.5cm}\includegraphics{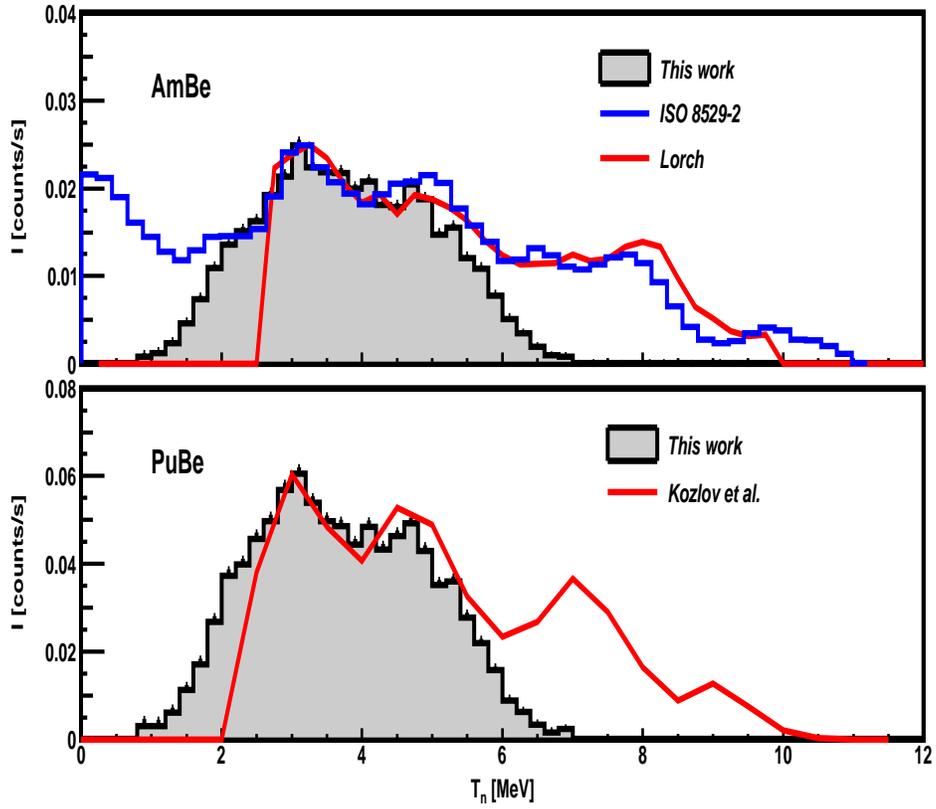}}
\caption{\label{figure:neutron_spectra}
Tagged-neutron results. The gray-shaded histograms are our 
livetime-corrected tagged-neutron measurements. Note the different 
y-axis scales in the top and bottom panel. Top panel: Am/Be. The 
full-energy spectrum (red line) of Lorch~\cite{lorch73} is also 
shown together with the Am/Be ISO 8529-2 reference neutron radiation 
spectrum. Bottom panel: Pu/Be. The full-energy spectrum (red line) 
of Kozlov et al.~\cite{kozlov68} is also shown.
(For interpretation of the references to color in this figure caption,
the reader is referred to the web version of this article.)
}
\end{center}
\end{figure}

Figure~\ref{figure:neutron_spectra} shows our actinide/Be tagged-neutron 
results obtained by measuring the neutron TOF between the NE-213 trigger 
and YAP(Ce) 4.44~MeV gamma-ray detectors. Our data are livetime-corrected 
yields -- they have not been corrected for neutron-detection efficiency 
or detector acceptance. Note that other than physically swapping the 
actinide/Be sources, absolutely no changes were made to our apparatus 
during data acquisition. In both panels, the maximum values of the Lorch
spectra at $\sim$3~MeV have been normalized to our distributions. Due to the 
neutron-tagging procedure, our data show no strength above $\sim$7~MeV as 
4.44~MeV is taken by the creation of the de-excitation gamma-ray. Our results 
for Am/Be presented in the top panel are shown together with the widely quoted
full-energy neutron spectrum of Lorch~\cite{lorch73}\footnote{
The results we present in this paper are for a newly acquired Am/Be
source that is different from the source that we used to produce the results
presented in Ref.~\cite{scherzinger15}. Due to flight-path differences
and the resulting resolution effects, it is difficult to exactly compare 
the two data sets. Nevertheless, over the energy range 2 -- 6 MeV, the 
difference in the tagged-neutron yields (normalized at 3~MeV) obtained with 
the two different sources is less than 2\%.
}
and the ISO~8529-2 reference neutron radiation spectrum. Agreement 
between the Lorch data and the reference spectrum is good between 
2.5~MeV and 10~MeV. Lorch did not observe the strength above 10~MeV 
present in the reference spectrum. The Lorch data display a sharp cutoff 
at 2.5~MeV which is attributed to an analysis threshold cut. The 
reference spectrum shows considerable strength below 2.5~MeV. Our data 
also show some strength in this region. Recall that our hardware threshold 
was 200~keV$_{ee}$ corresponding to a neutron energy of $\sim$1~MeV, and that 
no analysis threshold cut was employed. The agreement between our data, those 
of Lorch, and the reference spectrum in the region of overlap is excellent. 
The interested reader is directed to Ref.~\cite{scherzinger15} for a 
detailed discussion. Our results for Pu/Be presented in the bottom panel 
are shown together with those of Kozlov et al.~\cite{kozlov68}. The 
Kozlov et al.  measurement involved detecting the neutrons emitted from 
a series of ``homemade" Pu/Be sources in a cylindrical (30~mm long 
$\times$ 30~mm $\diameter$) stilbene crystal. Statistical uncertainties 
in these results were reported to be better than 3\% below 5~MeV and 
better than 20\% at 10~MeV. Similar to the Lorch data discussed above, 
the Kozlov et al. data also display a sharp cutoff at 2.5~MeV. While it
is not possible to determine the reason for this cutoff from the 
information presented in Ref.~\cite{kozlov68}, we again attribute it to 
an analysis threshold cut.  Agreement between the Kozlov et al. data 
and ours is good between 2.5~MeV and 6~MeV.  Finally, we note the
qualitative agreement between our Am/Be and Pu/Be results, the Lorch
and Kozlov et al. results, and the ISO ISO8529-2 reference neutron
radiation spectrum.

\begin{figure} 
\begin{center}
\resizebox{12.5cm}{7cm}{\hspace{-1.5cm}\includegraphics{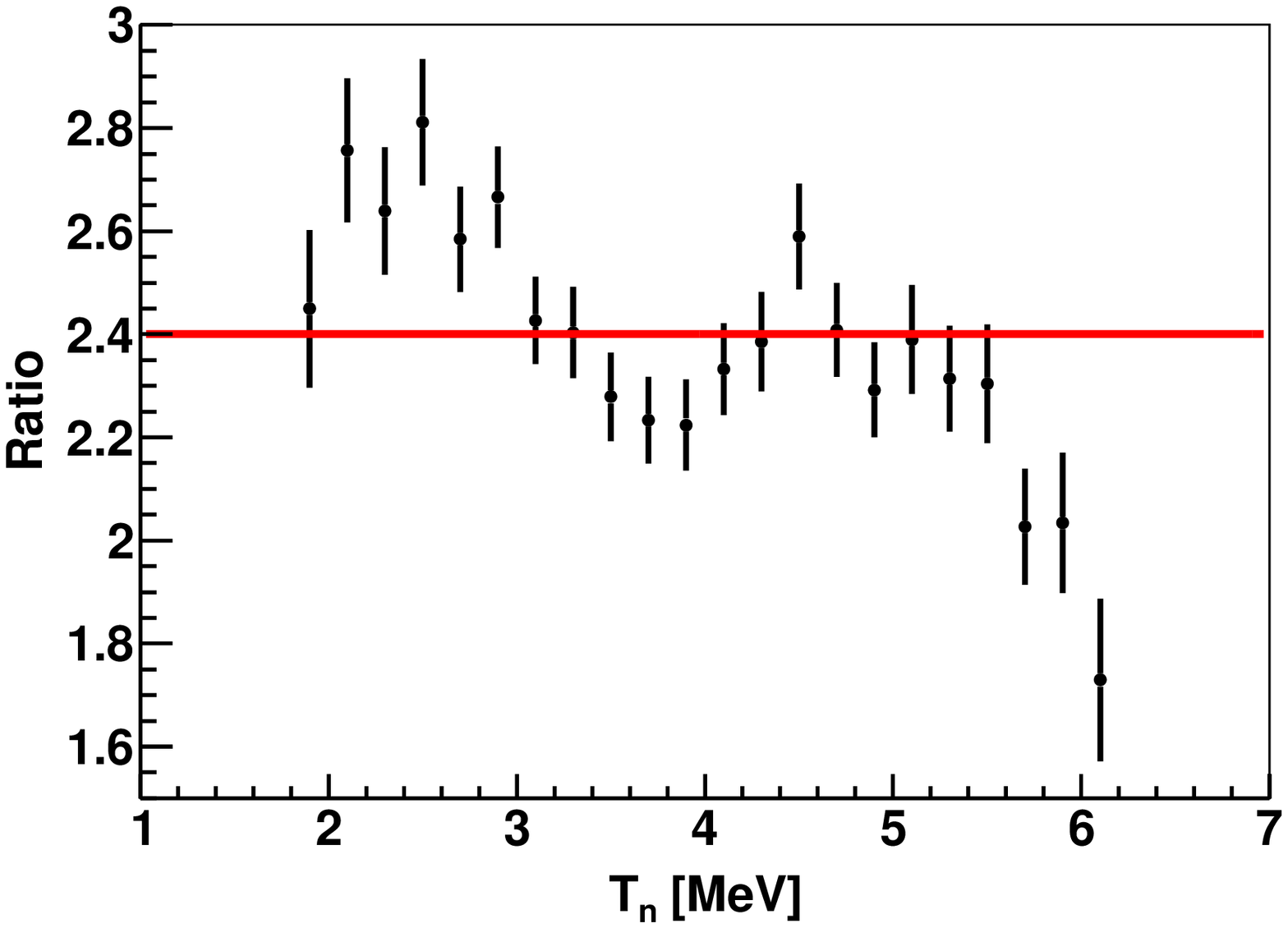}}
\caption{\label{figure:ambe_and_pube}
Livetime-corrected (Pu/Be) : (Am/Be) ratio of tagged-neutron yields.
The red line shows the average value.  (For interpretation of the 
references to color in this figure caption, the reader is referred to 
the web version of this article.)
}
\end{center}
\end{figure}

Figure~\ref{figure:ambe_and_pube} presents a quantitative comparison
between the livetime-corrected tagged-neutron yields for our Am/Be 
and Pu/Be sources. The (Pu/Be) : (Am/Be) ratio is taken from the 
gray-shaded regions presented in Fig.~\ref{figure:neutron_spectra}
and is displayed for the energy region over which we are confident 
that we are not simply seeing effects from applied hardware thresholds. The 
ratio of the integrals between 2 and 6~MeV is 2.4, which is in 
agreement with the original full-spectrum source-calibration measurements 
performed at the time of their acquisition. Explanation of the observed
fluctuation in the ratio as a function of neutron energy requires a more 
detailed study of the possible effects of source composition, neutron-detection
efficiency, and detector acceptances. This will be covered in future work.

\section{Summary}
\label{section:summary}

We have measured untagged gamma-ray and tagged-neutron yields from $^{241}$AmBe
and $^{238}$PuBe sources. Our untagged gamma-ray distributions ranged from 
1 -- 5~MeV. They were performed in an open environment with the same 
experimental conditions for both sources. The shapes of the distributions are 
very similar and are clearly dominated by the 4.44~MeV gamma-ray associated 
with the de-excitation of the first-excited state in $^{12}$C. We determined 
that our Pu/Be source emits about 2.6 times as many 4.44~MeV gamma-rays as our 
Am/Be source. Our tagged-neutron distributions ranged from 2 -- 6 MeV. They 
were performed in a shielded environment with the same experimental conditions 
for both sources. The shapes of the distributions are quite similar and agree 
well with previous data obtained for the respective sources. Further, the shape 
of the Am/Be distribution agrees well with the ISO 8529-2 reference neutron 
radiation spectrum. We determined that our Pu/Be source emits roughly 2.4 times 
as many taggable neutrons as our Am/Be source, also in reasonable agreement with 
the original full-spectrum source-calibration measurements performed at the time
of their acquisition. Observed differences in the details of the shape of the
neutron spectra will be the subject of future investigations.

\section*{Acknowledgements}
\label{acknowledgements}

We acknowledge the support of the UK Science and Technology Facilities 
Council (Grant nos. STFC 57071/1 and STFC 50727/1) and the European 
Union Horizon 2020 BrightnESS Project, Proposal ID 676548.

\newpage

\bibliographystyle{elsarticle-num}

\end{document}